\documentclass[aps,prl,twocolumn,superscriptaddress,showpacs]{revtex4}
\input{epsf}
\def\gsim{\mathrel{\rlap{\lower4pt\hbox{\hskip1pt$\sim$}}
    \raise1pt\hbox{$>$}}} 

\usepackage{epsfig}

\begin{document}
\title{Dynamics of entanglement in quantum computers with imperfections}
\author{Simone Montangero}
\email{monta@sns.it}
\affiliation{NEST-INFM $\&$ Scuola Normale Superiore, Piazza dei Cavalieri 7,
56126 Pisa, Italy}
\author{Giuliano Benenti}
\email{giuliano.benenti@uninsubria.it}
\homepage{http://www.unico.it/~dysco}
\affiliation{Center for Nonlinear and Complex Systems, Universit\`a degli
Studi dell'Insubria and INFM, Unit\`a di Como, Via Valleggio 11, 
22100 Como, Italy}  
\author{Rosario Fazio} 
\affiliation{NEST-INFM $\&$ Scuola Normale Superiore, Piazza dei Cavalieri 7,
56126 Pisa, Italy}

\date{\today}

\begin{abstract} 
The dynamics of the pairwise entanglement in a qubit lattice in the 
presence of static imperfections exhibits different regimes.
We show that there is a transition from a perturbative region, where 
the entanglement is stable against imperfections, to the ergodic regime, 
in which a pair of qubits becomes entangled with the rest of the lattice 
and the pairwise entanglement drops to zero. 
The transition is almost independent of the size of the quantum computer. 
We consider both the case of an initial maximally entangled and separable 
state. In this last case there is a broad crossover region in which the 
computer imperfections can be used to create a significant amount of 
pairwise entanglement. 
\end{abstract} 

\maketitle


Any practical implementation of a quantum computer will have 
to face errors, due to the inevitable coupling to the environment 
\cite{zurek,palma} or to device imperfections \cite{GS}. The effect of 
the environment is to introduce decoherence which sets the time scale 
over which quantum computation is no longer possible. The presence of 
static imperfections, although not leading to any decoherence, may 
be also decremental for the implementation of any quantum computational task. 
A small inaccuracy in the coupling constants induces errors in 
gating or an unwanted time evolution in the case in which the 
Hamiltonian cannot be switched exactly to zero.  If the  
imperfection strength increases, new phenomena occur and above a certain 
threshold the core of the computer can even ``melt'' 
due to the setting in of chaotic behavior~\cite{GS}. Previous investigations 
of the stability of quantum information processing in the presence 
of such effects mainly studied the fidelity of the quantum evolution as an 
indicator of the quality of the computation~\cite{zurek2,GS,simone}. 
In particular, in Ref.~\cite{GS} the fidelity was used to measure 
the stability of the  quantum memory, that is of a state loaded on a quantum 
computer with imperfections. 
A  more complete characterization of the stability of a quantum computation 
requires a deeper investigation of the state  of the system. Fidelity 
is one of the ways to characterize it. In this Letter we discuss the  
behavior of entanglement on approaching the transition to quantum chaos.
Entanglement is not only one of the most intriguing features 
predicted by the quantum theory but also a fundamental resource for 
quantum computation and communication~\cite{chuang}. Therefore 
studies of the stability of entanglement under decoherence and imperfection 
effects are, in our opinion, of interest.   

We model the quantum computer as a lattice of interacting spins
(qubits) where,  
due to the unavoidable presence of imperfections, the spacing between
the up and  
down states (external field) and the couplings between the 
qubits (exchange interactions) are both random. The 
entanglement properties in spin systems has recently attracted 
attention and we refer to~\cite{spinent} for a more detailed 
introduction. In this work we consider, for the first time, the effect 
of disorder in the couplings.

We consider $n$ qubits on a two-dimensional lattice, described by
the Hamiltonian
\begin{equation}
H=\sum_{i=1}^{n} \Delta_i \sigma_{i}^z + \sum_{i<j=1}^{n} 
J_{ij} \left( \frac{1+\gamma}{2} \sigma_{i}^x \sigma_{j}^x + 
\frac{1-\gamma}{2} \sigma_{i}^y \sigma_{j}^y \right),
\label{hamil}
\end{equation}
where the $\sigma_{i}^\alpha$ ($\alpha=x,y,z$) are the Pauli matrices 
for the qubit $i$ and the second sum runs over nearest-neighbor qubit 
pairs on the lattice. 
The energy spacing between the up and down states of a qubit is given by
$\Delta_i=\Delta_0+\delta_i$, where the $\delta_i$'s are 
uniformly distributed in the interval $[-\delta /2,\delta /2]$.
The parameter $\delta$ gives the width of the
$\Delta_i$ (single qubit energy) distribution around the average 
value $\Delta_0$. 
The couplings $J_{ij}$ are also uniformly distributed in the
interval $[-J,J]$. We consider $0\leq \gamma \leq 1$. For $\gamma=1$
Eq.~(\ref{hamil}) reduces to the disordered Ising model, while 
for $\gamma=0$ it gives the disordered XY model. For $\gamma=1$ 
the model defined by Eq.~(\ref{hamil})  has been proposed 
by Georgeot and Shepelyansky \cite{GS} to describe the hardware of 
a quantum computer, in which system imperfections generate 
unwanted interqubit couplings and energy fluctuations. 
For $J, \delta=0$ the spectrum of the Hamiltonian is composed of $n+1$ 
degenerate levels, and the interlevel spacing is $2\Delta_0$, which corresponds
to the energy required to flip a single qubit. 
We study the case $\delta,J \ll \Delta_0$, in which 
the degeneracies are resolved and the spectrum is composed by $n+1$ bands.
Since $\delta,J\ll \Delta_0$,
the interband couplings in (\ref{hamil}) can be neglected and every single 
band can be studied separately. We concentrate our studies on the central
band with zero magnetization ($\sum \sigma_i^z=0$). 
We assume free boundary conditions and set the energy scale $\Delta_0=1$. 

In this Letter, we investigate the evolution in time of the 
entanglement between two nearest neighbor qubits in the lattice. 
We confine our interest to the entanglement 
of formation~\cite{bennett} between two spins in the lattice. 
The propagation of  entanglement in disordered-free systems 
has been analyzed in Ref. \cite{amico}. In the same spirit we 
study the entanglement evolution for an initially maximally 
entangled pair of qubits: 
\begin{equation}
|\Psi_B(0)\rangle = 
\frac{1}{\sqrt{2}} (|01\rangle + |10\rangle) \otimes 
| 0101\dots01 \rangle \; .
\label{bells}
\end{equation}
We will also analyze the generation of the entanglement, due to 
interqubit coupling, of two spins which are initially in a separable
state. In this case we start from the wave vector  
\begin{equation}
|\Psi_S(0) \rangle= |01\rangle \otimes  
| 0101\dots01 \rangle \; .
\label{seps}
\end{equation}
In both cases, the initial state, Eqs.~(\ref{bells}-\ref{seps}), 
evolves according to 
the Hamiltonian defined in Eq.~(\ref{hamil}). We compute the reduced
density matrix 
$\rho_{12}(t)$, obtained from the wave function $|\Psi(t)\rangle$
after all the spins except those at sites 1 and 2 have been 
traced out: $\rho_{12} (t) = {\rm Tr}_{3,...,n} 
|\Psi(t) \rangle\langle \Psi(t)|$. We choose qubits 1 and 2 to be
nearest neighbors on the border of the lattice (their position  
in the lattice is not crucial for the results here reported). 
Their state at time $t$, described by the density matrix
$\rho_{12}(t)$, is mixed since qubits 1 and 2 become entangled with 
the rest of the lattice. 
We quantify the entanglement by means of the concurrence~\cite{bennett},
defined as   
$C=\max\{\lambda_1-\lambda_2-\lambda_3-\lambda_4,0\}$,
where the $\lambda_i$'s are the square roots of the eigenvalues 
of the matrix $R=\rho_{12}\tilde{\rho}_{12}$, in descending
order; the spin-flipped density matrix is defined by 
$\tilde{\rho}_{12}=(\sigma_1^y\otimes\sigma_2^y) 
\rho_{12}^\star (\sigma_1^y\otimes \sigma_2^y)$ (in 
this definition the standard basis $\{|00\rangle,|01\rangle,
|10\rangle,|11\rangle\}$ must be used).
Another dynamical quantity  widely used to characterize 
the stability of a state under perturbations is the fidelity \cite{GS}. 
It is defined as $f(t)=|\langle \Psi(0) | \Psi(t)\rangle|^2$,
where $|\Psi(t)\rangle=\exp(-iHt)|\psi(0)\rangle$ (we set $\hbar=1$).
For weak interqubit coupling $f(t)$ is close to 1 for all times. 
For strong coupling we enter the so-called 
quantum chaos regime, and the fidelity essentially drops to zero \cite{GS}.
In the following we contrast the behavior of the fidelity with 
that of the concurrence. 
In addition to the different regimes found in the analysis of the 
fidelity, the concurrence gives additional information related to 
the choice of the initial state.

\begin{figure} 
\centerline{\epsfxsize=8cm \epsffile{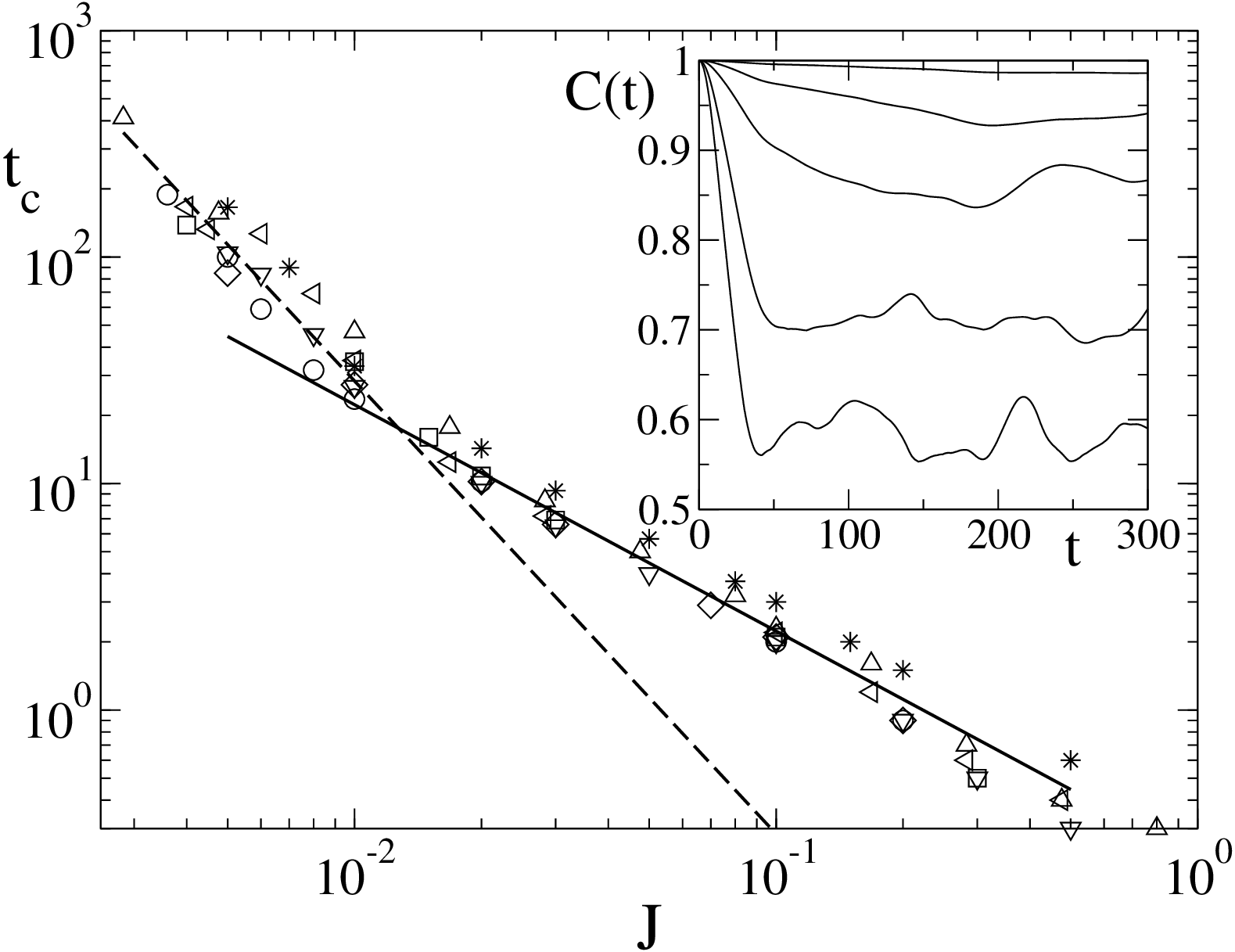}}
\caption{Concurrence time scale $t_c$ as a function of $J$,
for $\gamma=1$, $\delta=0.2$, initial state $|\Psi_B(0)\rangle$, 
and $n=4$ (triangles up), $6$ (triangles left), $8$ (triangles down), 
$10$ (circles), $12$ (squares), and $14$ (diamonds)
qubits. Stars represent $n=10$ and $\gamma=0$.  
The full and the dashed lines are proportional to $J^{-1}$ and 
$J^{-2}$, respectively. 
Inset: Concurrence as a function of time for $n=10$ and $\gamma=1$. 
From top to bottom: $J= 2,5 \times 10^{-3}; 1,2,3 \times  10^{-2}$.
To smooth statistical fluctuations, here and in the following 
figures data are averaged over $N_r$ different random 
configurations of the imperfections $\delta_i$ and $J_{ij}$: 
$N_r=50$ for $n=4,6,8,10$ and $N_r=30,20$ for $n=
12,14$, respectively.}
\label{fig1}       
\end{figure} 

{\it Initially entangled state -~} 
We first consider the time evolution of $C(t)$ starting 
from the (maximally entangled) initial state $|\Psi_B(0)\rangle$. 
If the coupling strength $J$ is weak, then the time evolution of $C(t)$ 
can be obtained by means of perturbation theory in the coupling $J$.
Only the quantum register states 
directly coupled to the  initial state are involved
(the quantum register states are the eigenstates 
of Hamiltonian (\ref{hamil}) at $J=0$:
$|\alpha\rangle=|\alpha_1,\alpha_2,...,\alpha_n\rangle$, with
$\alpha_i=0,1$). 
Since the exchange  interaction has a two-body nature, 
the matrix elements are zero when the state binary strings 
differ in more than two digits. 
Furthermore, only nearest neighbor qubits interact, 
and therefore to first order in perturbation theory 
only order of $n$ quantum register states are involved. 
It is important to note that the coupling to a state which 
differs from the initial state $|\Psi_B(0)\rangle$ in the 
polarization state of two qubits other that 1 and 2 does not affect  
the pairwise entanglement between qubits $1$ and $2$. 
To first order in the perturbation theory, these two qubits 
become entangled only with nearest neighbor qubits. Therefore the 
concurrence can be described in terms of superposition 
between a small number of quantum register states, independently 
of the total number $n$ of qubits.  
In the perturbative regime, the frequencies of these oscillations are 
$\sim \delta$, around an average value $ \sim 1 - (J/\delta)^2$. 
Perturbation theory breaks down when the typical interaction 
matrix element $J$ between directly coupled states 
becomes of the order of their energy separation $\Delta E
\sim \delta/n$. This follows from the fact that 
two quantum register states are directly coupled when two qubits 
are flipped. This changes the energy by $\sim \delta$, and there are
the order of 
$n$ states directly coupled inside this energy interval. 
Therefore the perturbative regime
is limited to $J<J_p\sim \delta/n$ \cite{GS}. 
Above this threshold, it is known \cite{flambaum,GS} that the 
fidelity decays exponentially: 
$f(t)\approx \exp(-\Gamma_f t)$, with the rate $\Gamma_f\sim J^2 \rho_f$
determined by the Fermi golden rule, where $\rho_f=1/\Delta E
\sim n/\delta$ is the density of directly coupled states.
The transitions involving
qubits other than 1 and 2 do not change the concurrence.  
Therefore the density of directly coupled states relevant for
this quantity is given by ${\rho}_c\sim 1/\delta$.
Hence we expect the concurrence to decay exponentially, 
$C(t)\sim \exp(-\Gamma_c t)$, with the rate $\Gamma_c\sim
J^2 {\rho}_c\sim J^2/\delta$. This expectation is 
borne out by the numerical data of Fig.~1, which show that 
the concurrence time scale $t_c^F$, defined by the condition 
$C(t_c^F)=0.96$, is inversely proportional to $J^{2}$
(the result does not depend on the value chosen).  
\begin{figure} 
\centerline{\epsfxsize=8cm \epsffile{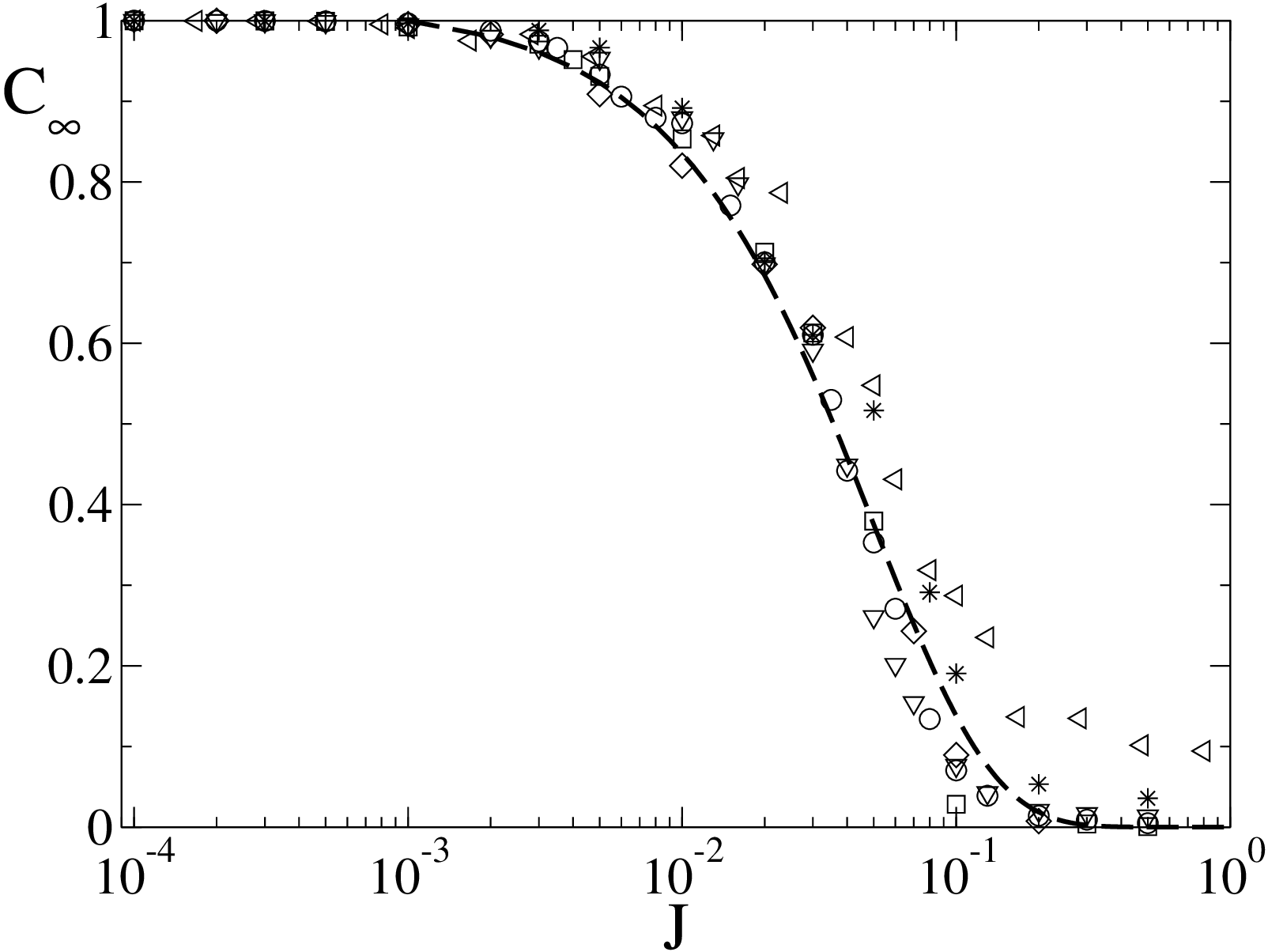}}
\caption{Concurrence saturation value $C_\infty$ as a function 
of $J$ with same parameter values and meaning of the symbols 
as in Fig.~1.
The dashed curve represents the fit 
$C_\infty (J) = \exp(-A (J-J_0))$, with $A\approx 20 $
and $J_0 \approx 1.0\times 10^{-3}$.
\label{fig2}}       
\end{figure} 
In the regime of strong coupling, $J \gsim \delta$, the time evolution
of the fidelity is given by~\cite{flambaum,GS}
$f(t)\approx\prod_{i<j}\cos(J_{ij} t)$,  
where the product runs over nearest neighbor qubits.
Thus the fidelity decay is Gaussian: 
$f(t)\approx\exp(-\sum_{i<j} J_{ij}^2 t^2)\sim \exp(-J^2 n t^2)$.
In computing the concurrence decay, we must again take into 
account that the rotations of qubits different from 1 and 2 do not
change their pairwise entanglement. Therefore we expect 
$C(t)\sim \exp(-J^2 t^2)$. This gives a concurrence time scale
$t_c^E\propto J^{-1}$, in good agreement with the 
numerical data of Fig.~1.
Since in the Fermi golden rule regime the concurrence time scale 
$t_c^F\sim \delta/J^2$, while in the ergodic regime  
$t_c^E\sim 1/J$, the transition between these two regimes 
takes place at $J=J_E\sim \delta$ (we checked numerically
that $J_E\propto \delta$).  
The data of Fig.~1, taken at different number
of qubits, seem to collapse on a single curve, in agreement
with our theoretical expectations.

It is also interesting to study the saturation value of the concurrence, 
$C_{\infty}=(\lim_{t \to \infty} \int_{t-\tau}^t C(t') dt')/\tau$.
In order to smoothen the fluctuations, we average our numerical data 
over both different random configurations of $\delta_i$ and $J_{ij}$
and a large enough  
time interval $\tau$ (we take $\tau$ equal to $1/10$ of the 
total integration time $t$, with $t$ sufficiently long to observe the  
saturation of $C(t)$).  
In Fig.~2, we study $C_\infty$ as a function of the coupling strength $J$, 
starting from the entangled state $|\Psi_B(0)\rangle$. Again we can 
distinguish three regimes. In the perturbative regime $J<J_p \sim \delta/n$
the saturation value is close to one (more precisely,
$1-C_\infty\propto (J/\delta)^2$).  
In the ergodic regime at $J>J_E\sim \delta$ the concurrence 
$C_\infty$ is equal to zero. Quantum chaos is characterized by the 
ergodicity of the eigenfunctions $|\phi_i\rangle$ 
($i=1,...,N=2^n$) of Hamiltonian (\ref{hamil}). 
This means that, by expanding $|\phi_i\rangle$ 
over the quantum register states, $|\phi_i\rangle = 
\sum_{\alpha} c_{\alpha} |{\alpha}\rangle$, the coefficient $c_{\alpha}$
are randomly fluctuating and have amplitudes
$|c_{\alpha}|\sim 1/\sqrt{N}$. In this regime, the 
complexity of the eigenstates, characterized by the entropy 
$S_i=-\sum_{\alpha} |c_{\alpha}|^2\log |c_{\alpha}|^2$,
is maximal, that is $S_i \approx n$. We note that indeed the maximum 
entropy criterion $S_i\approx n$ gives the ergodicity threshold 
$J_E\sim \delta$~\cite{silvestrov}. Due to quantum ergodicity, 
for times $t>>t_c^E\sim 1/J$ the wave function $|\Psi(t)\rangle$
is a random superposition of the quantum register states:
$|\Psi(t)\rangle=\sum_\alpha a_\alpha (t)|\alpha\rangle$, 
where the coefficients $a_\alpha(t)$ have amplitudes 
$\sim 1/\sqrt{N}$ and random phases. The diagonal 
elements of the reduced density matrix $\rho_{12}$ are given 
by $[\rho_{12}(t)]_{\alpha_1,\alpha_2;\alpha_1,\alpha_2}=
\sum_{\alpha_3,...,\alpha_n}|a_{\alpha_1,...,\alpha_n}(t)|^2$. 
The value of these elements is $\approx 1/4$, since they are 
the sum of $N/4$ positive terms, and the value of each term is 
$\sim 1/N$. The off diagonal matrix elements of 
$\rho_{12}$ are instead given by the sum of $N/4$ terms with random 
signs, and therefore their value is $\sim 1/\sqrt{N}$.
For such a reduced density matrix the concurrence is equal to zero. 
Physically this means that in the ergodic regime the entanglement 
is multipartite and shared between all the qubits, and therefore the 
pairwise entanglement drops to zero. In this regime the coupling to 
the rest of the system acts as a dephasing channel, and the reduced density 
matrix becomes essentially diagonal, thus destroying the pairwise
entanglement.   

\begin{figure} 
\centerline{\epsfxsize=8cm \epsffile{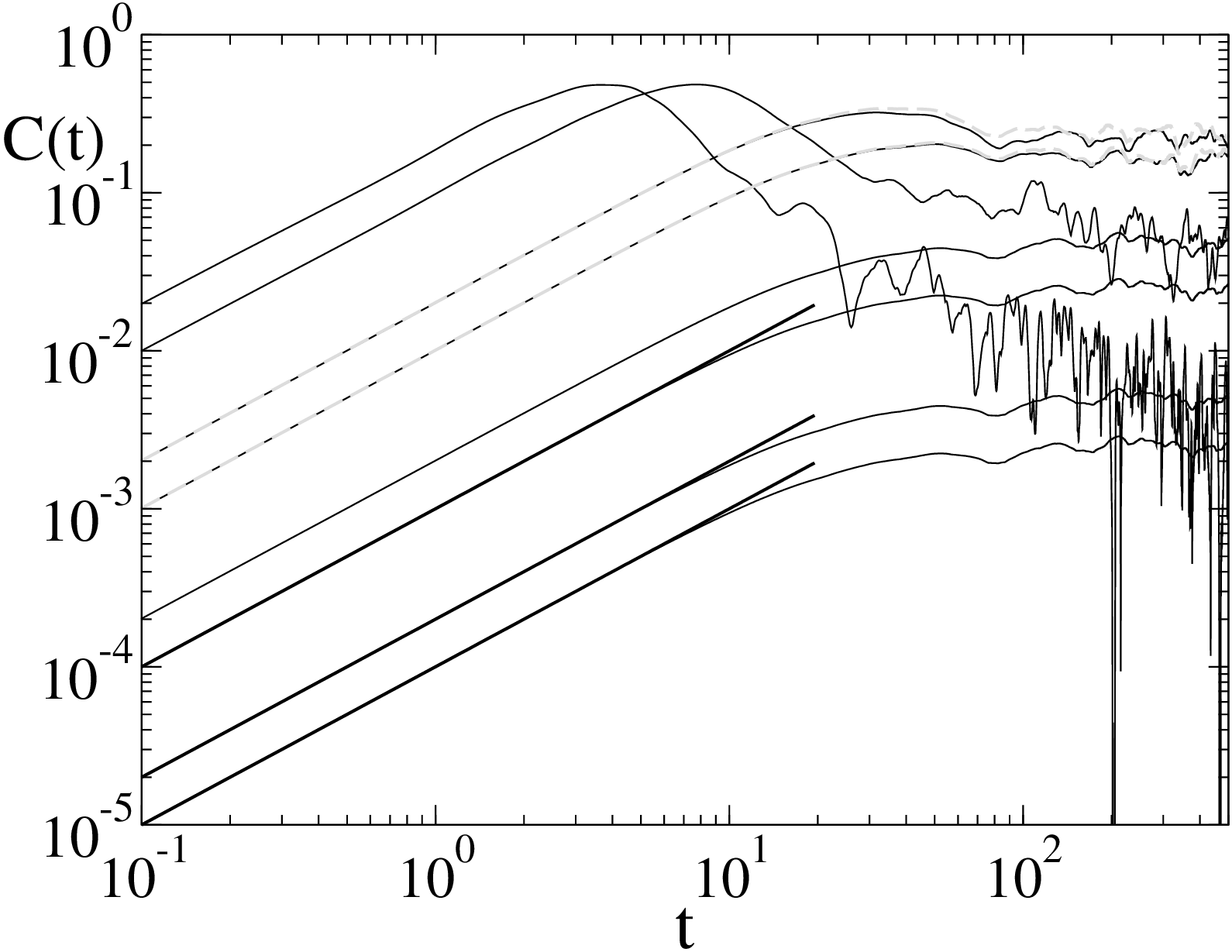}}
\caption{Concurrence $C(t)$ for $\gamma=1$ (full curves) and 
$\gamma=0$ (dashed gray curve), $\delta=0.2$, initial state
$|\Psi_S(0)\rangle$, $n=10$, and, from bottom to top, $J= 1,2 \times
10^{-4}; 1,2 \times 10^{-3}; 1,2 \times 10^{-2}; 1,2 \times
10^{-1}$. The straight lines give the time dependence $C(t)= J t$
predicted by the perturbation theory.
\label{fig3}}       
\end{figure} 

{\it Initially separable state -~} 
Since the Hamiltonian in Eq.~(\ref{hamil}) can create entanglement due to the 
exchange term between neighboring qubits~\cite{amico}, it is interesting to 
consider also separable states as initial states. 
In the perturbative regime $J<J_p$, the concurrence oscillates
with frequency $|\delta_1 -\delta_2|$ 
between 0 and a maximum value 
$2 |J_{12}|/|\delta_1 -\delta_2|$. At times $t\ll 1/|\delta_1 -\delta_2|$, the 
perturbation theory gives  $C(t)\approx 2 |J_{12}| t$, 
in agreement with the numerical data of Fig.~3
(after disorder averaging we get $C(t)=Jt$). In the ergodic regime 
$J>J_E$, a maximum concurrence of the order of one is 
generated after a time $ \sim 1/J$. After this time 
the entanglement of the pair rapidly drops to zero.  
In the intermediate regime the concurrence reaches its maximum 
and then saturates to a value which is nearly independent of $J$. Therefore 
the generation of pairwise entanglement
is an interesting characteristic of this region.

\begin{figure} 
\centerline{\epsfxsize=8cm \epsffile{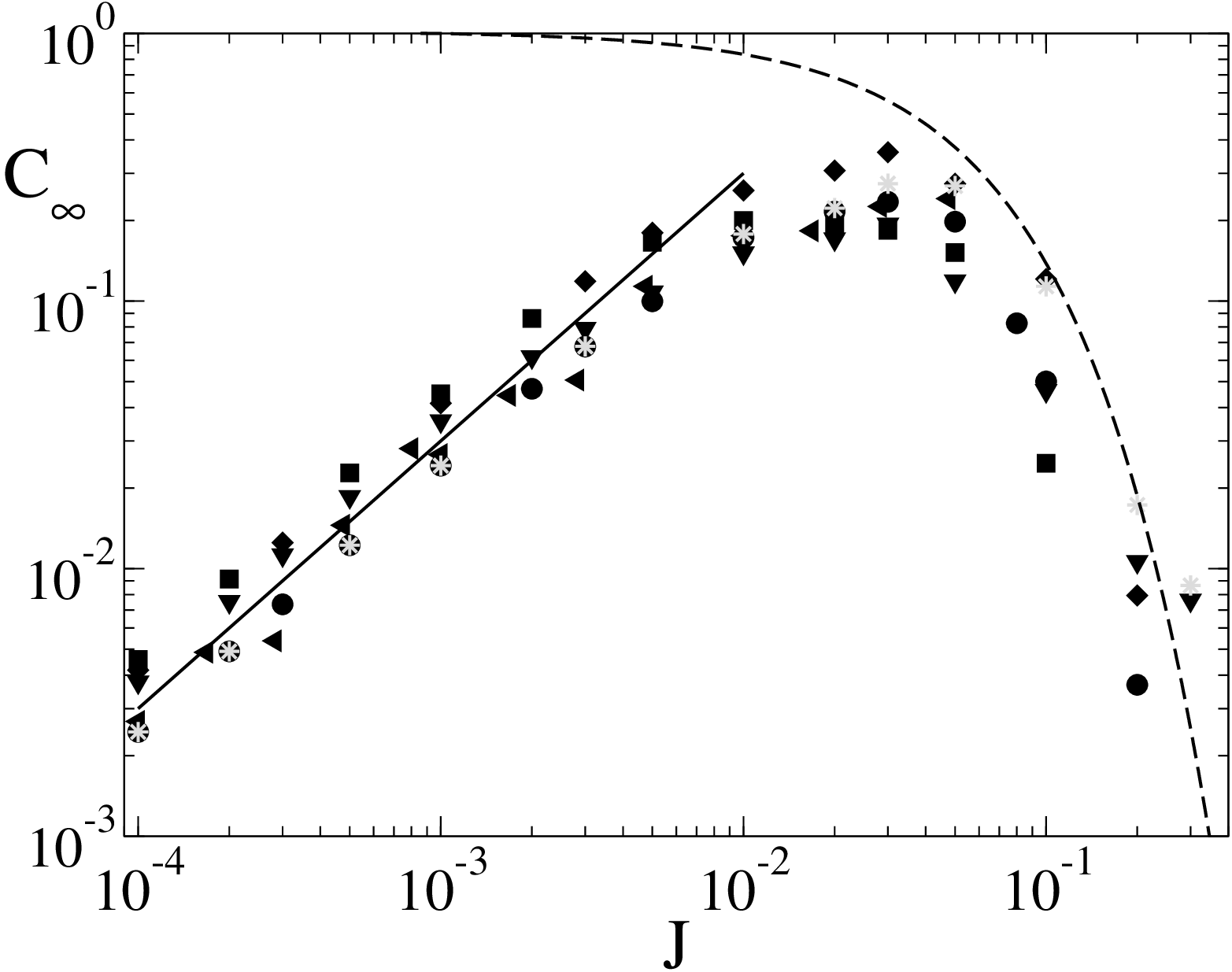}}
\caption{Concurrence saturation value $C_\infty$ as a function of $J$ 
for $\gamma=0$ (black symbols) and $\gamma=1$ (gray stars),
$\delta=0.2$, initial state $|\Psi_S \rangle$ (filled symbols). Symbols
have the same meaning as in Fig.~1. The straight line is proportional to $J$.
The dashed line is the same as in Fig.~2.
\label{fig4}}       
\end{figure}

The saturation value $C_{\infty}$ as a function of the coupling 
$J$ is shown in Fig.~4. In the perturbative regime
$C_\infty\propto J/\delta$ whereas in the ergodic regime
$C_\infty=0$. It is interesting to notice that there is a 
broad region in which the interactions create a 
significant amount of entanglement, quantified by a 
concurrence $C_\infty\approx 0.2-0.3$. 
We also note that in the ergodic regime $J>J_E$ the saturation 
value $C_\infty=0$ does not depend on the initial conditions, 
as expected. On the contrary, when $J<J_E$ the concurrence 
saturation value depends on the initial state vector, implying
the absence of ergodicity. 

As it can be seen from our data shown 
in Figs.~1-4, we did not found any significant
change in the dynamics of entanglement as a function of $\gamma$. 
Even though a further symmetry is added to the system when 
$\gamma=0$ (the invariance with respect to rotations about 
the $z$-axis), this has little influence on entanglement. 

In summary, we have studied the dynamics of the pairwise 
entanglement in a qubit lattice in the presence of static 
imperfections, and characterized three different regimes:
(i) the perturbative regime, in which the entanglement is stable 
against imperfections, (ii) the crossover regime, in which the 
imperfections degrade the concurrence of an initially 
entangled pair but can also drive a significant entanglement 
generation, and (iii) the ergodic regime, in which a pair of 
qubits becomes entangled with the rest of the lattice and 
therefore the concurrence of the pair drops to zero. 
We stress two important points of our findings from the 
point of view of quantum computation. First of all, 
the pairwise entanglement is destroyed above a coupling strength 
which is independent of the size of the quantum computer.
Moreover, there is a broad crossover region in which the 
computer imperfections can be used to create a significant 
amount of pairwise entanglement. Finally we want to emphasize that 
our analysis leads to the conclusion that the spin system behaves as  
a dephasing environment only in the ergodic regime.
In the integrable limit, even an infinite system is unable to 
dephase completely the initial state. We believe that this observation 
may be relevant in understanding the behavior of spin or more complex 
baths.

We acknowledge very fruitful discussions with L. Amico, A. Osterloh, 
and F. Plastina. This work was supported by the 
EU (IST-SQUBIT), RTN2-2001-00440, HPRN-CT-2002-00144.

\newpage


\end{document}